\documentclass[twocolumn]{aastex63}
\usepackage{amssymb,amsmath}
\usepackage{graphicx}
\usepackage{xcolor,natbib}
\usepackage{multirow}
\usepackage[T1]{fontenc}
\usepackage{ae,aecompl}
\usepackage{newtxtext, newtxmath}
\usepackage{float}
\usepackage{subfigure}
\usepackage{longtable}
\usepackage{enumitem}
\usepackage{longtable,listings}
\usepackage[flushleft]{threeparttable}
\usepackage{parcolumns}
\def\kms{${\rm km\,s}^{-1}$}

\submitjournal{ApJ}

\shorttitle{Velocity Offset of SDSS J1557}
\shortauthors{Zheng et al.}

\begin{document}

\title{Velocity offset between emission and absorption lines might be an effective indicator of dual core system}

\correspondingauthor{Xueguang Zhang; Qirong Yuan}
\email{xgzhang@njnu.edu.cn; yuanqirong@njnu.edu.cn}

\author{Qi Zheng}
\affiliation{School of Physics and Technology, Nanjing Normal University, No. 1,	Wenyuan Road, Nanjing, 210023, P. R. China}

\author{Shuang Liu$^{3}$}
\affiliation{Purple Mountain Observatory, Chinese Academy of Sciences, 10 Yuanhua Road, Nanjing, Jiangsu 210023, China\\
$^{3}$School of Astronomy and Space Sciences, University of Science and Technology of China, Hefei, Anhui 230026, China\\
}

\author{Xueguang Zhang$^{*}$}
\affiliation{School of Physics and Technology, Nanjing Normal University, No. 1,	Wenyuan Road, Nanjing, 210023, P. R. China}

\author{Qirong Yuan$^{*}$}
\affiliation{School of Physics and Technology, Nanjing Normal University, No. 1,	Wenyuan Road, Nanjing, 210023, P. R. China}

\begin{abstract}
This paper presents a detection of significant velocity offset between emission and absorption lines for a dual core system in 
SDSS~J155708.82+273518.74 (= SDSS~J1557).
The photometric image of SDSS~J1557 exhibits clear two cores
with a projected separation of $\sim$2.2 arcseconds (4.9 kpc) determined by $\sc{GALFIT}$.
Based on the applications of the commonly accepted pPXF code with 636 theoretical SSP templates, the host galaxy contribution  
can be well determined.
Then, the emission line features of SDSS~J1557 can be well measured
after subtraction of host starlight.
It is found that the velocity offset of emission lines with respect to absorption lines reaches $458 \pm 13$ \kms.
According to the Baldwin-Phillips-Terlevich (BPT) diagram, SDSS J1557 is a composite galaxy.
In addition, SDSS J1557 can well fit the $M_{\rm BH}-\sigma_{\ast}$ relation of bulges and the galaxy merger would not change this relation.
Two reasonable models (say, AGN-driven outflow vs. dual core system) have been discussed to explain this velocity offset.
 The model of AGN-driven outflow fails to interpret the systematic redshift of emission lines and similar velocity offsets for various emission lines in SDSS~J1557.
 A significant velocity offset between emission and absorption lines might be an effective indicator of dual core system.

\end{abstract}


\section{Introduction}
Galaxy merger is an inevitable stage in the evolution of almost all galaxies \citep{Be80,Si98,Zh04, Me06,Ma10,Ma13,Ma19,Zu22} and this merger would result two massive black holes merging into a more massive black hole.
Supermassive black holes (SMBHs), which can be found in the center of nearly all galaxies \citep{Ko95,War21}, are closely related to their host galaxies.
At the stage where the two black holes are separated from kpc scale to pc scale, they can be observed as dual core systems such as AGN-AGN pairs, AGN-galaxy pairs, galaxy-galaxy pairs \citep{Liu16,St16}.
Then dual AGN (AGN-AGN pairs) will evolve into a binary black hole (BBH) system with sub-pc scale separation \citep{De19,Ko20}.
Since the BBH may survive for around a Hubble time in galaxies \citep{Yu02},  BBH systems should not a rare phenomenon. However, due to the small physical separation \citep{Ba08} and misclassification in morphological merger identification \citep{De07,Lo08}, there are not enough conclusive BBH systems in galaxies.
%Finally, BBH will be gradually merged into a more massive black hole.

In astronomy, it is always a hot topic to find and identify the AGN pairs  with different scales \citep{Bo09,Ro09,Sm10,Sh10,Er12,Co13,Gr15,Li16,Ba17,Wa17,Ko20,Zxg21}.
The BBH system can be used to explain the phenomena of long-standing quasi-periodic oscillations (QPOs) \citep{Gu14,Gr15Na,Br20,Yang21}, double-peaked broad emission lines \citep{Sh10,Ki201,Te22} and special 
X-shaped radio galaxies \citep{Me02,Li04,Zxg07}.
%In the BBH system, when the accretion disk of the primary black hole (BH) is crossed by the secondary BH in an eccentric orbit \citep{Ri10}, a signal of periodicity would be found.
Dual AGN system can result in double-peaked narrow emission lines \citep{Co091,Ge12,Co14} and/or two cores in photometric images \citep{Zh04,Kh17,Go19,Zh21} .
%\citep{La05, Gi16,Na17} or 
When two black holes are separated with kpc scale, dual AGNs would produce double-peaked profiles as the effect of redshift and blueshift towards observers.	
%X-shaped radio galaxies show the main lobes and a pair of wings, which bend in opposite directions \citep{Gi16}.%
%NGC~6240 is a typical example with two compact X-ray cores, which could hint the existence of BBH \citep{Ko03}.%
In this paper, the focus is to report a new dual core system with no double-peaked emission profiles but large offset in radial velocity between emission and absorption lines.

In center of galaxy, it is conclusive that the SMBH is coincident with the dynamical center \citep{Re20}.
It is generally believed that similar situation also holds for other galaxies, but a few effects can temporarily make the SMBH stay away from this equilibrium position \citep{Pe21}.
Considering the shallow galactic potential, in the case of low galaxy mass, SMBHs could never settle at the galaxy center \citep{Be19,Re20}.
In the case of higher galaxy mass, galaxy mergers are most likely responsible for the SMBH motions.
Throughout the process of merger, SMBHs and their host galaxies will move relatively and cause spatial offsets.
At the same time, each SMBH of the galaxy would be offset in space and velocity from the merging center.
It is possible that active galaxies with velocity offset \citep{Co14} or spatially resolved pairs of AGNs \citep{Ge07} can be observed.

AGN possesses a compact broad-line region (BLR) with the size of about 1 pc and an extended narrow-line region (NLR) with the size of about 1000 pc \citep{De22}.
If both of the SMBHs power the AGN, at the case of sufficient gas, the two nuclei could be visible when their NLRs do not overlap.
In this scenario, the merger-remnant galaxy spectrum shows two sets of emission lines and both the two sets of emission lines would be separated from each other and from the stellar continuum light of host galaxy in space and  velocity \citep{Be18,Do20,Ki20}.
Those emission lines include the strong [O~{\sc iii}] emission line, which indicates AGN activity.
Such merger-remnant galaxies with two distinct sets of emission lines are referred to as dual AGNs.	
For example, type 2 quasar SDSS J1048+0055 had two discrete sources and double peak narrow emission lines in [O~{\sc iii}] $\lambda\lambda$4959\AA, 5007\AA, which was considered as a dual AGN \citep{Zh04}.

If only one of the merging SMBHs powers the AGN, the merger-remnant galaxy spectrum also would exhibit only single-peaked AGN emission lines.
Meanwhile, the set of emission lines is separated from the stellar continuum light of host galaxy in space and velocity.
Such merger-remnant galaxies with only one set of emission lines are referred to as offset AGNs.	
In addition, when the two nuclei stay at very small separation, any anisotropy of the radiative linear momentum will result in a gravitational-wave recoil \citep{Fi83,Re89,Ma22}, and the signatures could be gained by pulsar timing arrays \citep{Br21}.
Depending on mass ratios, spin orientations and spin magnitudes, the merging black holes are predicted to attain a kick velocity between a few hundred and a few thousand \kms  \citep{Ca07,Sc07,Lo11,Ki17}.
Many candidates for observational recoil have been proposed, but they all have their own complications \citep{Ko12,Bl16}.

The inspiralling SMBHs during the galaxy merger could cause the double-peaked emission lines and velocity-offset emission lines.
The double-peaked narrow emission lines have been used as a common method of selecting dual core candidates \citep{Wa09,Wo14,Li18}.
However, there are several studies to suggest that the presence of double-peaked narrow emissions is not necessarily always associated with interacted dual AGNs \citep{Sh11,Fu12,Ru19,Du22}.
Alternatively, single-peaked narrow emission lines, which show velocity offset relative to system, also can be used to identify the candidate of dual core system \citep{Co17}.
\citet{Ba08} reported that NGC~3341 has a blueshifted velocity of about 190\kms with 5.1 kpc projected distance away from the primary galaxy.
Based on the DEEP2 Galaxy Redshift Survey, \citet{Co09} found 30 AGNs with single-peaked velocity-offset [O~{\sc iii}] emission lines from 1881 red galaxies, and inspiralling SMBHs in merger-remnant galaxies can better explain this phenomenon than other models.
%Few months later, \citet{Co091} reported that galaxy COSMOS J100043.15+020637.2 ($z$=0.36) has two bright point sources near the center of galaxy and a spatial offset of 1.75$\pm$0.03 $h^{-1}$ kpc with Hubble Space Telescope imaging, a 150$\pm$40 \kms velocity offset between two AGNs.
%\citet{Do09} pointed out that SDSS J092712.65+294344.0 possibly hosts a massive black hole binary with the mass ratio about 0.3 and has the velocity offset of 2650 \kms in broad emission lines, and this may be caused by the motion of gas orbiting around secondary black hole.%
After the search of the COSMOS survey, \citet{Ci10} reported that CXOC J100043.1+020637 ($z$=0.359) has two compact optical sources with a distance of $0.^{\prime \prime}495\pm 0.^{\prime \prime}005$, the velocity offset between the broad and narrow H$\beta$ emission lines is 1200 \kms, and both gravitational wave recoiling black hole and a triple black hole system could explain this phenomenon.
\citet{Co13} found five offset AGN candidates from 173 type 2 AGNs ($z$<0.37) in Galaxy Evolution Survey, and pointed out that the AGNs with velocity-offset emission lines would increase with redshift because the galaxy mergers fraction increases with redshift.
Based on data of the Sloan Digital Sky Survey (SDSS) DR7, \citet{Co14} reported that 4\% - 8\% type 2 AGNs ($z$<0.21) exhibit velocity offset when considering projection effects in observed velocities and those offset AGNs with higher luminosity are preferentially explained by galaxy mergers.
%\citet{Pe21} reported that J0437+2456 has a recession velocity of 4910 \kms according to neutral hydrogen emission, which is differs from 4810 \kms H$_{2}$O megamaser velocity of central SMBH, but the specific nature of the SMBH remains unclear.%

Here, we present a galaxy, SDSS J155708.82+273518.74 ($z$=0.125), with two clear cores and significant velocity offset between emission and absorption lines.
Meanwhile, the emission lines are redshifted with respect to absorption lines. 
The two cores, velocity offset and redshift of emission lines
make SDSS J155708.82+273518.74 very special. 
The paper is organized as follows.
The acquisition of the object, possible physical models for this object are presented in Section 2. Future applications of velocity-offset emission lines on searching dual core systems are prospected in Section 3.
In Section 4, a summary of main findings is given.
In this paper, we have adopted the cosmological
parameters of $H_{0}=70{\rm km\,s}^{-1}{\rm Mpc}^{-1}$,
$\Omega_{\Lambda}=0.7$ and $\Omega_{\rm m}=0.3$.

\section{Main Results and Discussions}
\begin{figure*}
\centering\includegraphics[width=16cm,height=5.3cm]{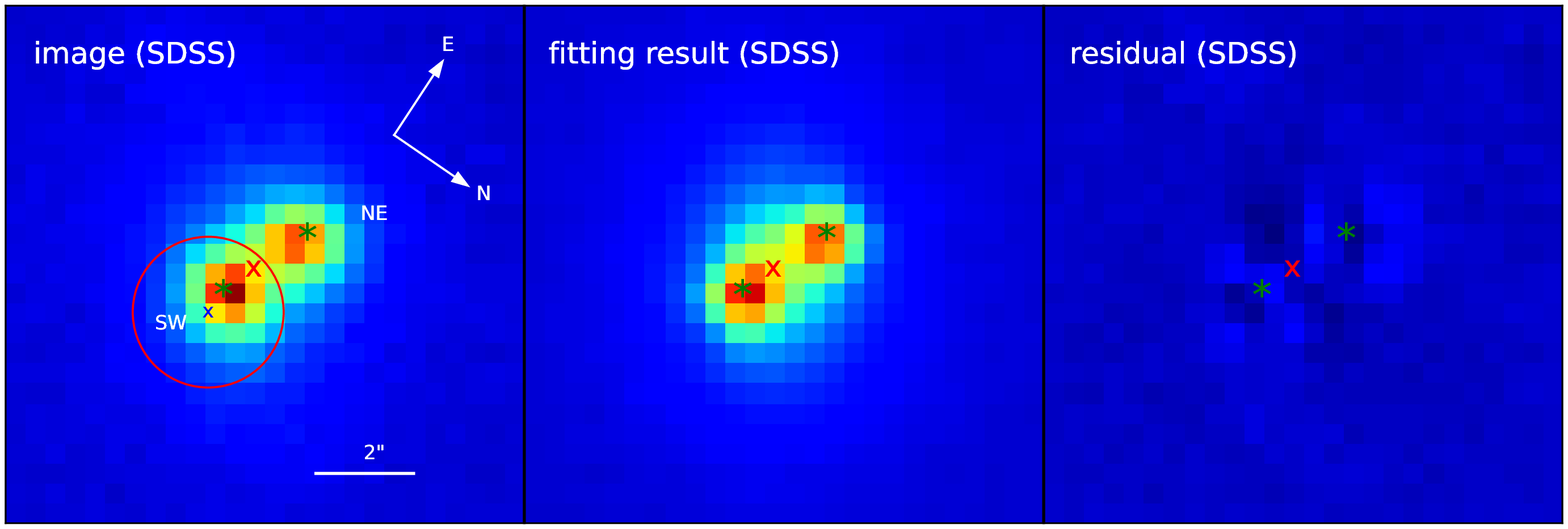}
\centering\includegraphics[width=16cm,height=5.3cm]{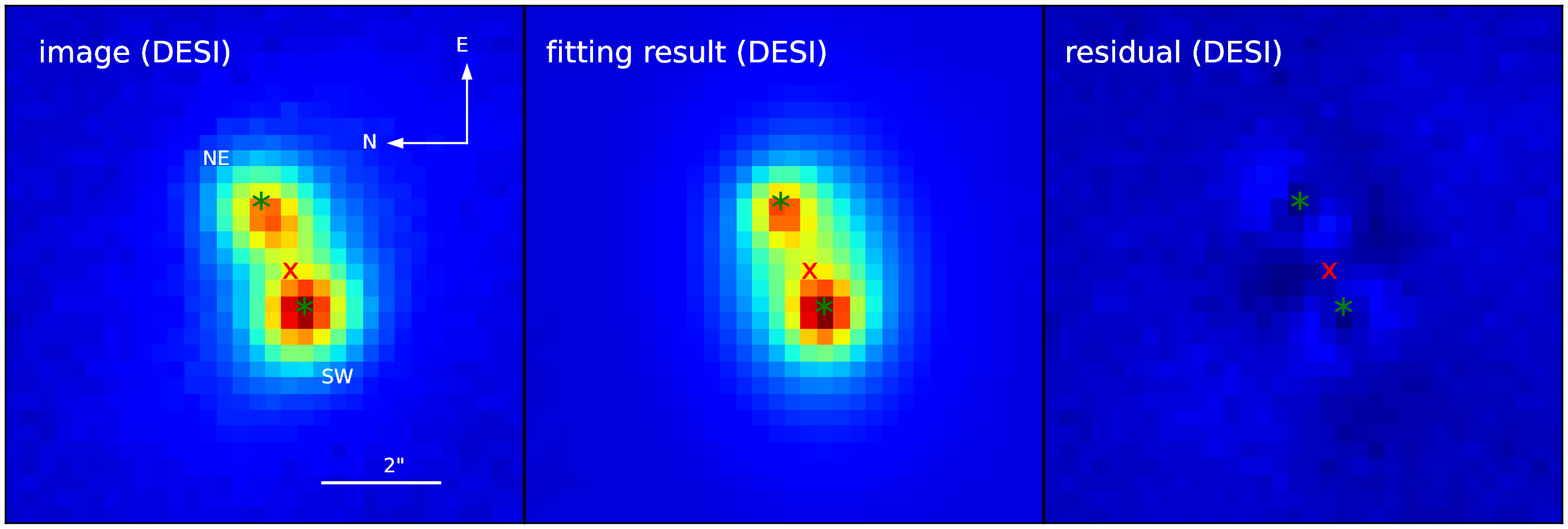}
\caption{Top three panels show results from the $\sc{GALFIT}$ analysis: image of SDSS J1557 in r-band from SDSS (left panel); best fitting result of $\sc{GALFIT}$ model (middle panle); and residuals(right panel).
Bottom three panels show corresponding results of SDSS J1557 in z-band from DESI.
In each panel, the center of host galaxy and two cores determined by $\sc{GALFIT}$ are represented in the red cross and the green snowflakes, respectively.	
Information of photometric data source of each panel is shown in white characters.
The orientation (in white arrow symbols) and plotting scale of 2 arcseconds are shown in white, and the regions of the two bright cores are marked as NE and SW in white characters in left two panels. In top left panel the red circle represents the drilled fiber (3\arcsec~in diameter) of SDSS J1557 spectrum and the blue cross shows the center of drilled fiber.
}
\label{fig8}
\end{figure*}

In order to find dual core system with obvious velocity offset (velocity offset>150 \kms), 
we select galaxy in the Sloan Digital Sky Survey (SDSS) DR16 \citep{Ah20} with SQL (Structured Query Language) Search tool (\url{http://skyserver.sdss.org/dr16/en/tools/search/sql.aspx}).
The applied SQL query in detail is as follows:
\begin{lstlisting}
  SELECT MJD, PLATE, FIBERID, Z, SNMEDIAN,
  V_off_balmer ,V_off_forbidden
  FROM SpecObjAll AS S
  JOIN GalSpecLine AS L
  ON S.specobjid = L.specobjid
  WHERE S.class ='galaxy' and
  (S.subclass ='starforming' or
  S.subclass ='starburst') and
  S.SNMEDIAN>15 and
  S.Z < 0.35 and S.zwarning = 0 and
  abs(V_off_balmer-V_off_forbidden)<30 and
  ((abs(V_off_balmer)>150 and
  V_off_balmer>5*V_off_balmer_err) or
  (abs(V_off_forbidden)>150
  and V_off_forbidden>5*V_off_forbidden_err))
\end{lstlisting}
The explanation of selection criteria above are shown as follows:
\begin{itemize} 
\item The application of "S.class ='galaxy' and (S.subclass ='starforming' or
S.subclass ='starburst')" is to choose quiescent galaxies to avoid the effects of AGN-driven outflows.	
\item The application of "S.SNMEDIAN>15" is to select SDSS galaxies with high quality spectra with signal-to-noise  larger than 15, in order to find more reliable emission line parameters.
\item The application of "S.Z < 0.35 and S.zwarning = 0" is to find SDSS galaxies with redshift smaller than 0.35 to ensure H$\alpha$ emission lines included in SDSS spectra.
\item The application of "abs(V\_off\_balmer-V\_off\_forbidden)<30 and ((abs(V\_off\_balmer)>150 and V\_off\_balmer>5*V\_off\_balmer\_err) or (abs(V\_off\_forbidden)>150
and V\_off\_forbidden>
5*V\_off\_forbidden\_err))" is to ensure obvious offset between absorption lines and emission lines.
\end{itemize}
As a result, two galaxies (Plate-MJD-Fiber: 1392-52822-400 and 2603-54479-395) are collected and we only choose the one galaxy with two unambiguous cores in photometric image (Plate-MJD-Fiber: 1392-52822-400) as shown in top left panel of Figure\ref{fig8}.
The peculiar source is SDSS J1557 (RA:239.286769; Dec:27.588540) with a redshift of 0.125.
In addition, in order to get accurate information, we recalculate the velocity offset of SDSS J1557 in the later paragraph rather than use the information in the SDSS provided public database of galSpecLine (\url{http://skyserver.sdss.org/dr16/en/help/docs/tabledesc.aspx}).

To analysis the source positions of SDSS J1557 quantitatively,
$\sc{GALFIT}$ \citep{Pe02,Pe10} is applied.
Here, the general S{\'e}rsic profile \citet{Se63} is chosen to describe host galaxy properties:
\begin{equation}
I(r)=I_e\exp[-b_n((\frac{r}{r_e})^{\frac{1}{n}}-1)].
\end{equation}
Here, $r_{\rm e}$ is the effective radius enclosing 50\% of the light along major axis, $I_{\rm e}$ is the pixel surface brightness at effective radius $r_{\rm e}$, $n$ is the S{\'e}rsic index, and $b_{\rm n}$ is the normalization constant.
We select the five nearest stars in $r$-band image to construct the point-spread function (PSF) with $\sc{EPSFBuilder}$ \citep{An00,An16} function in python. 
In the fitting procedure, two PSF models for the two cores and a flat background sky are applied.
Only one S{\'e}rsic model for host galaxy is applied here
because a late-stage merger can be well-represented by a single, partially-coalesced galaxy that encompasses the two nuclei.
The fitting result is shown in top panels of Figure \ref{fig8} and the positions of the two cores in the photometric image of SDSS are (RA=15:57:08.85; DEC=27:35:18.84) and (RA=15:57:08.99; DEC=27:35:19.54).
The separation between the two bright cores in SDSS is 2.2 arcseconds, corresponding to a projection distance of about 4.9 kpc.
Similarly, due to the better spatial resolution (0.27 arcsecond/pixel better than 0.396 arcsecond/pixel in SDSS) of the Dark Energy Spectroscopic Instrument (DESI) legacy imaging surveys \citep{De19}, we also fit the photometric image of SDSS J1557 in z-band of DESI with $\sc{GALFIT}$, as shown in bottom panels of Figure \ref{fig8}.
The determined coordinates of the two cores in DESI are (RA=15:57:08.85; DEC=27:35:18.84) and (RA=15:57:08.99; DEC=27:35:19.54), respectively. And the separation between the two bright cores in SDSS is 2.1 arcseconds, which is similar to the result from SDSS.

\begin{figure}
	\centering\includegraphics[width=5cm,height=8cm,angle=90]{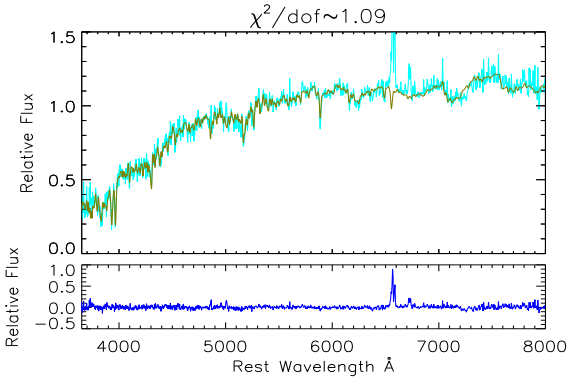}
	\caption{The SDSS spectrum of SDSS J1557 and the starlight determined by the pPXF method.
		In the top panel, the solid cyan line denotes the spectrum of SDSS J1557, and the solid dark yellow line shows pPXF-method-determined starlight. In the bottom panel, the blue line shows the line spectrum calculated by the SDSS spectrum minus the starlight.
	}
	\label{fig2}
\end{figure}

It is generally accepted that the dual core system SDSS~J1557 may generate
special spectrum.
Due to strong contribution from host galaxy, it is necessary to determine the starlight from the host galaxy.
For measuring and calibrating host galaxy composition in the SDSS AGN spectra, there are several similar Simple Stellar Population (SSP) methods \citep{Br03,Ka03,Ci05,Zxg19} based on stellar population template spectrum, such as principal component analysis (PCA) \citep{St09}, independent component analysis (ICA) \citep{Kr04}, and penalized pixel fitting (pPXF) code  \citep{Ca17}.
Here, pPXF code is adopted to calibrate the starlight included in the spectrum.
The pPXF method is originally described in \citet{Ca04}, which takes a maximum penalized likelihood method in the pixel space to fit the spectra, and the line-of-sight velocity distributions (LOSVD) are described by Gauss-Hermite parameterization \citep{Ge18}.
It is particularly useful to obtain kinematics from integral-field
spectroscopic data \citep{Ca04}, but there are also a number of applications for stellar population analyses \citep{Ca12,Sh15,Li17}.
The SSP templates can be chosen	in considerable freedom and
the best fitting from pPXF should be a weighted sum of SSPs, resulting in a composite spectral model. 	
For simplicity, the SSP templates can be generated from the empirical Medium resolution Isaac Newton Library of Empirical Spectra (MILES) stellar library \citep{Kn21}.
The 636 SSPs used in the paper are consisted of 53 population ages and 12 metallicities (from minus 2.27 to 0.4).
The population ages range from 0.03 Gyr for the youngest to 14 Gyr for the oldest.
%And then regularization is used to smooth out the expected star formation history.
The SSP-method-determined starlight is clearly shown in the top panel of Figure \ref{fig2}, and its corresponding residuals are presented in the bottom of Figure \ref{fig2}.
In addition, the spectral resolution of SDSS is 1500 at 3000 \AA~ and 2500 at 9000 \AA~ and the mean value for the spectral resolution is about 70 \kms \citep{Gr05}. Also descriptions on SDS spectral resolutions can be found in 
\url{http://www.sdss3.org/dr9/spectro/spectro_basics.php}. 
Through the head information of the zero extension of spec-1392-52822-0400.fits, drilled fiber position
is plug\_ra=239.28674 and plug\_dec=27.588527 (RA=15:57:08.82;DEC=27:35:18.70).
The photometric image of SDSS J1557 with the drilled fiber in red circle is shown in top left panel of Figure \ref{fig8} and the blue cross represents the center of the drilled fiber.
Considering that the narrow line region is about 3 kpc estimated by [O{\sc iii}] luminosity \citep{Ha13,Ha14}, the drilled fiber can totally cover the SW core and the narrow line region of the NE core.
Due to the coverage of narrow line region of the NE core, the center of the NE core out of the drilled fiber would not affect the contributions of the NE core for the emission lines.
However, it is difficult to correct effects of aperture loss on intrinsic emission line properties due to loss of multiple spectra observed with different aperture.
But the aperture loss mainly have effects on the flux rather than dynamic properties of emission lines.
The main objective of this paper is to detect and report a dual core system through properties of velocity offset between absorption lines and emission lines, so effects of aperture loss do not change our final conclusion.
Meanwhile, simiply considering the similar effects on flux, the results on BPT diagram, which will be discussed in the paper, would not be changed due to similar flux ratios of different emission lines.

\begin{table*}
	\caption{Features of main emission lines}
	\begin{center}
		\begin{tabular}{c|c|c|c|c}
			\hline\hline
			Emission lines	& Center wavelength (\AA)  & Line flux ($10^{-17}{\rm erg/s/cm^{2}}$)& $\sigma$ (\AA) & Velocity offset (\kms) \\ \hline
		$\ast$	H$\beta$	& 4863.72   & 32.24$\pm$5.46 & 3.09 & 457.77  \\ \hline
		$\ast$	[O\sc{iii}]	& 4960.15  & 14.69 & 4.56  & 375.58 \\ \hline
		[O\sc{iii}]	& 5008.09$\pm$0.90  & 44.07$\pm$8.80 & 4.60$\pm$0.94  & 384.57$\pm$53.91 \\ \hline	
		$\ast$[N\sc{ii}] 	& 6551.03 & 91.98$\pm$12.37 & 4.75 & 447.49 \\ \hline
		H$\alpha$	& 6566.02$\pm$0.28  & 241.14$\pm$15.24 & 4.17$\pm$0.27 & 457.91$\pm$12.80 \\ \hline	
		[N\sc{ii}] 	& 6586.45$\pm$0.42 & 147.45$\pm$13.74 & 4.78$\pm$0.45 & 446.91$\pm$19.14 \\ \hline
		[S\sc{ii}] 	& 6719.96$\pm$1.27  & 60.09$\pm$14.67 & 4.73$\pm$1.36 & 467.98$\pm$56.71 \\ \hline
		[S\sc{ii}] 	& 6733.87$\pm$1.22  & 39.39$\pm$12.80 & 3.41$\pm$1.19 & 446.93$\pm$54.37 \\ \hline
		\hline
		\end{tabular}
	    \begin{tablenotes}
		\item[*]The emission lines with $\ast$ are rather weak, which are given by tied with other emission lines and show large uncertainties.
	\end{tablenotes}
	\end{center}
	
\end{table*}

\begin{table*}
	\caption{List of separation and velocity offset for different dual core systems}
	\begin{center}
		\begin{tabular}{c|c|c|c}
			\hline\hline
			Source	& Distance between nuclei  & Velocity offset & Reference \\ \hline
			NGC 3341	& 5.1 kpc  & 190 \kms & \citep{Ba08}  \\ \hline
		%	COSMOS J100043.15+020637.2	& 2.586$\pm$0.047 kpc  & 150$\pm$40 \kms & \citep{Co091} \\ \hline
		%	CID-42	& 2.46$\pm$0.02 kpc  & 1200 \kms & \citep{Ci10}\\ \hline	
			NDWFS J143316.48+353259.3	& 15 kpc  & 100.9$\pm$9.1 \kms & \citep{Co13} \\ \hline
			NDWFS J143317.07+344912.0	& 16 kpc  & 143.0$\pm$17.3 \kms & \citep{Co13} \\ \hline
			SDSS J1557 & 4.9 kpc &   457.91$\pm$12.80 \kms  & this work \\ \hline
			\hline
		\end{tabular}
	\end{center}
	
\end{table*}

\begin{figure*}
	\centering\includegraphics[width=16cm,height=7cm]{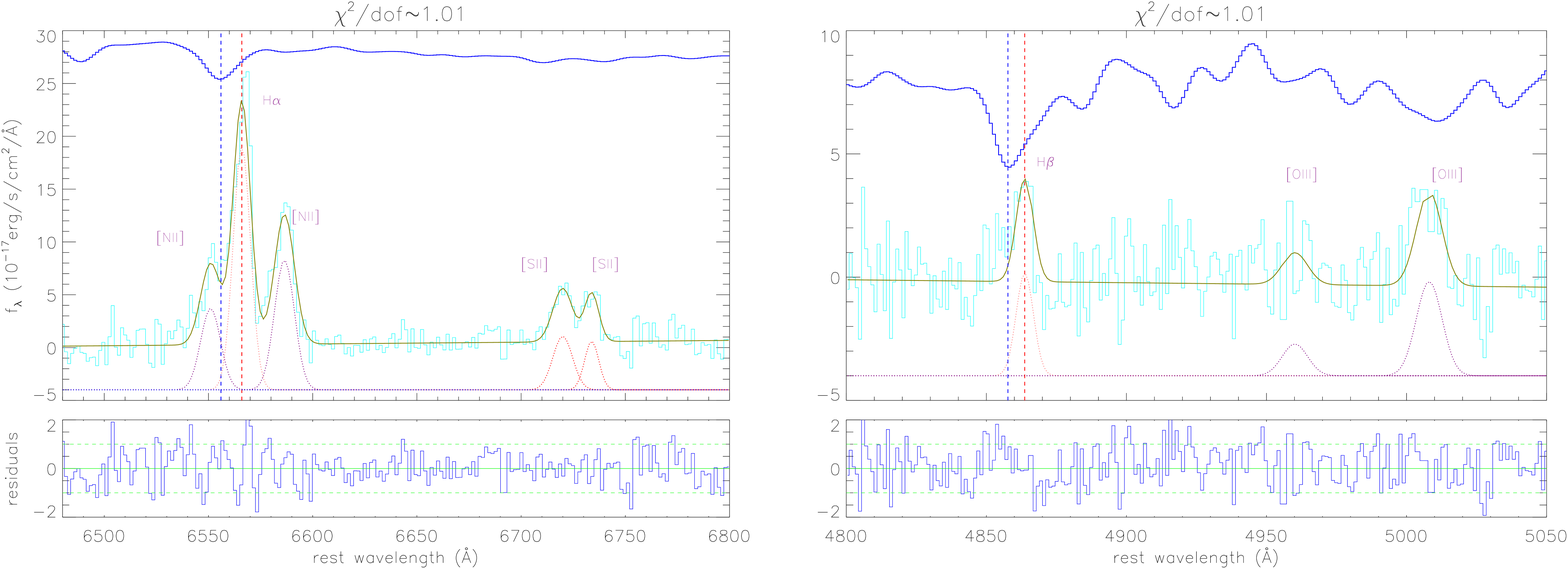}
	\caption{The best fitting results to the emission lines around H$\alpha$ emission line (left panel) and H$\beta$ emission line (right panel) by Gaussian functions.
	In each panel, the solid cyan line represents the 
	line spectrum after subtracting starlight determined by pPXF code from host galaxy and the solid dark yellow line represents the best-fitting results.
	The name of each emission line is shown in purple character.
	In the top of each panel, the solid blue line shows the relative absorption line profiles determined by pPXF code.
	The vertical dashed blue and red lines in the top of each panel represent the peak of absorption and emission lines, respectively.
	The solid blue line in the bottom of each panel represents the
	residuals calculated by the line spectrum minus the best fitting results and then divided by uncertainties of SDSS spectrum. 
	The horizental solid and dashed green lines show residuals=0,$\pm$1, respectively.
}
	\label{fig3}
\end{figure*}

\begin{figure}
	\centering\includegraphics[width = 8cm,height=6cm]{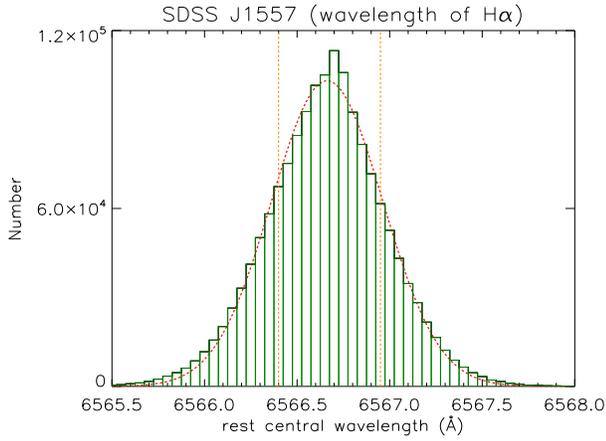}
	\caption{The MCMC technique determined posterior distribution of central wavelength of H$\alpha$ emission line. The dashed red line shows the Gaussian fitting of the posteriori distribution. The width of each green bin is 0.05. The two vertical dashed orange lines represent the marked values relative to 16\% and 84\% area covered below the values.}
	\label{fig00}
\end{figure}

The absorption lines of SDSS J1557 can be well determined by pPXF code.
Due to the strong H$\alpha$ emission line and absorption line,
the part of the spectrum close to Balmer lines is exhibited with Gaussian functions after deducting contribution from host galaxy.
Here, we mainly consider the emission lines around H$\alpha$ and H$\beta$ (rest wavelength from 4800 to 5050 \AA~and from 6450 to 6800 \AA) and tied the H$\alpha$ and H$\beta$ emission lines together.
The redshift and width of H$\beta$ emission line is fixed with H$\alpha$ emission line.
Two power law components are applied to describe possible
continuum emissions underneath the emission lines around H$\alpha$ and H$\beta$ emission lines, respectively.
In addition, Gaussian components of [N{\sc ii}] ([O{\sc iii}]) doublet have the same redshift and line width in velocity space.
Due to the low signal-to-noise ratio (SNR) and effects of spectral splicing, the flux of [O{\sc iii}] $\lambda$4959\AA~ is determined by the flux of [O{\sc iii}] $\lambda$5007\AA~ with the theoretical flux ratio 1:3.
Based on the Levenberg-Marquardt least-squares minimization
technique, the emission lines around H$\alpha$ and H$\beta$ emission lines can be well measured.
In order to test the reliability of excluding the broad component, two model functions are applied to describe the emission lines of SDSS J1557.
The first model function (model 1) is that each of the emission lines is described with one narrow plus one broad Gaussian functions ($\chi_1^2 \sim 393.47, dof_1=397$). 
The second model function (model 2) is that each of the emission lines is described with only one narrow Gaussian function ($\chi_2^2 \sim 417.93, dof_2=414$).
It is useful to determine whether single Gaussian function  is preferred 
through the log likelihood ratio test \citep{Pa221}.
The likelihood ratio, the Akaike information criterion (AIC)), can be evaluated as \citep{Ru191}:
\begin{equation}
	AIC=2N-2log{\mathcal L},
\end{equation}
where N is the number of parameters($N_1 =38$ and $N_2 =21$), $\mathcal{L}$ is the likelihood and $log\mathcal{L}=const-\chi ^2/2$. According to $\bigtriangleup AIC=AIC_1 -AIC_2\sim9.54$, model 2 is $exp(\bigtriangleup AIC/2)\sim118$ times more likely than model 1.
Therefore, each of the emission lines is described with single narrow Gaussian function.	
The best fitting results are shown as solid dark yellow line in Figure \ref{fig3} with
$\chi_1^2/dof_1=1.01$ (sum of squared residuals divided by degree of freedom). 
In addition, the uncertainties are the formal 1$\sigma$ errors computed from the covariance matrix for the final determined best-fit model parameters returned by the $\sc{MPFIT}$ function in IDL.
The 1$\sigma$ uncertainty reported by $\sc{MPFIT}$ tend to be underestimated when several parameters are tied together,
so we accurately estimate the reliability on central wavelength and line flux through the maximum Likelihood method combining with the Markov Chain Monte Carlo (MCMC) technique \citep{fh13}, which is implemented by $\sc{emcee}$ package in Python.
We describe emission lines around H$\alpha$ and H$\beta$ like the method above, but the H$\beta$ emission line do not tied with H$\alpha$ emission line.
The evenly prior distributions of model parameters of the emission lines are accepted with the following limitations, central wavelength of narrow emission lines within theoretical values $\pm20$\AA~,
second moment $\sigma\in[20,~800]$ in unit of ${\rm  km/s}$, flux $A\in[0,~500]$ in unit of $10^{-17}{\rm erg/s/cm^2}$.
Due to the importance of the central wavelength of H$\alpha$ for the velocity offset, we only show the MCMC technique determined posterior distribution of the central wavelength of H$\alpha$ here.
The MCMC technique determined posterior distribution of the central wavelength of H$\alpha$ is shown in Figure \ref{fig00} and the vertical dashed orange lines mark the values relative to 16\% and 84\% area covered below the values, respectively.
The MCMC technique determined 1$\sigma$ uncertainty 0.275\AA~ of the central wavelength of H$\alpha$ emission line is totally similar to the 1$\sigma$ uncertainty 0.28\AA~of the central wavelength of H$\alpha$ emission line determined by $\sc{MPFIT}$. Moreover, the MCMC technique determined uncertainties of parameters of the other narrow emission lines are also similar to uncertainties determined by $\sc{MPFIT}$.

The dashed blue and red lines in each panel of Figure \ref{fig3} represent the peak of absorption lines and emission lines, respectively.
Thus it can be determined that the velocity offset of the H$\alpha$ emission line relative to the absorption line in SDSS J1557 is $457.91 \pm 12.80$ \kms.
Due to higher quality of H$\alpha$  absorption line, the H$\alpha$ absorption line is chosen as a standard frame to measure the velocity offset of various emission lines, and the results are shown in table 1.
Here, the line parameters of each narrow emissions described by
Gaussian component are also presented in Table 1.

\begin{figure*}
	\centering\includegraphics[width = 8cm,height=6cm]{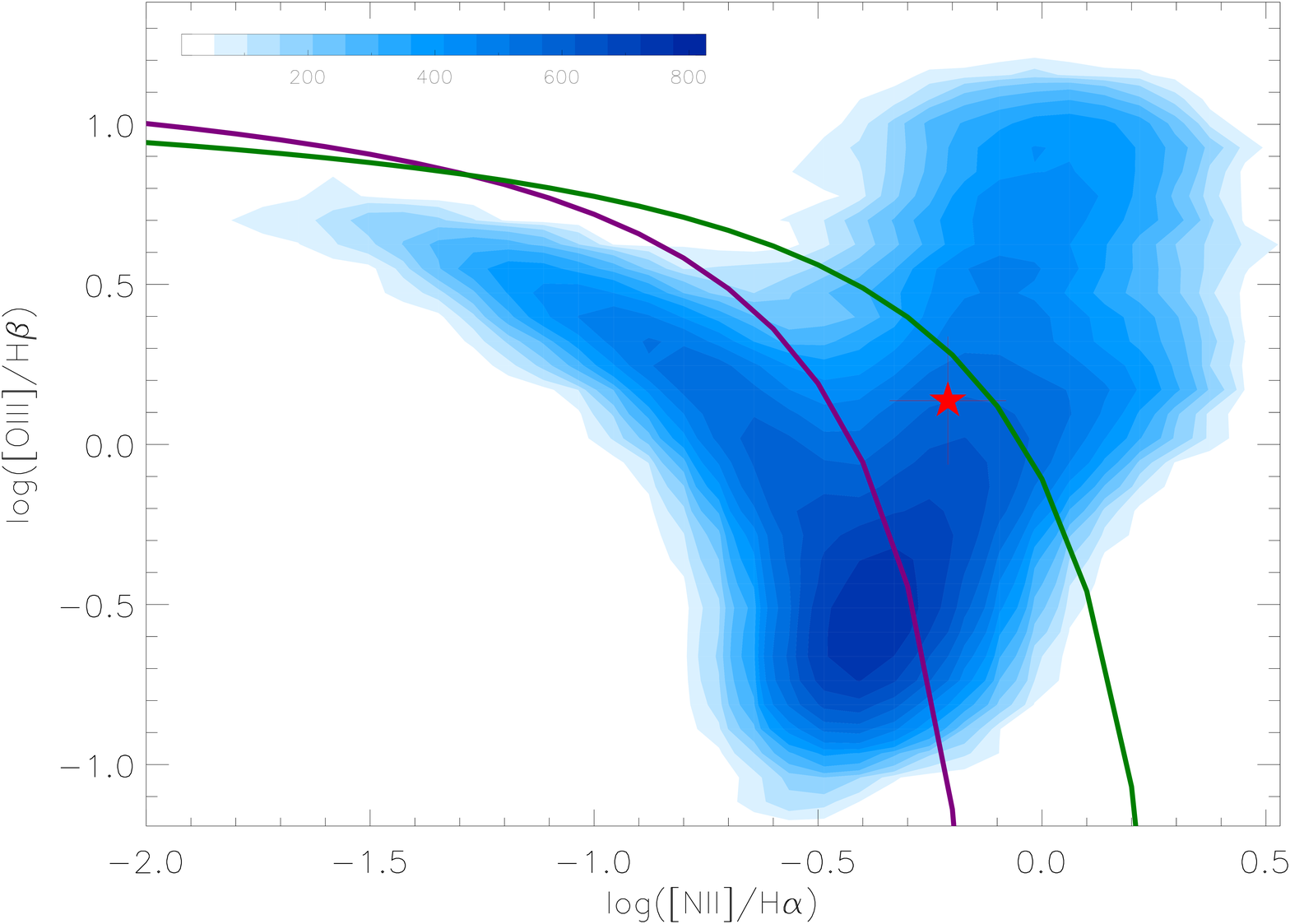}
	\centering\includegraphics[width = 8cm,height=6cm]{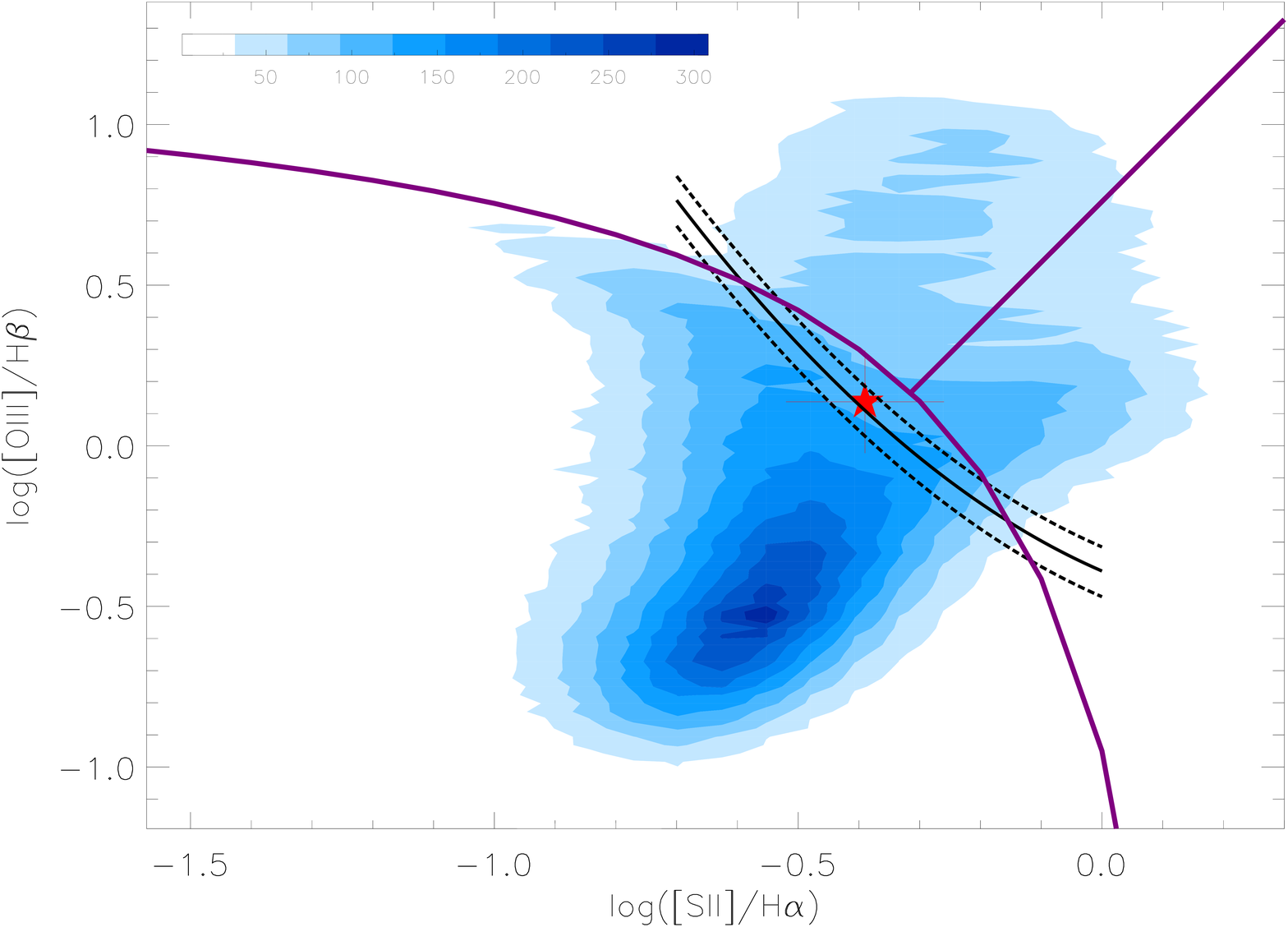}
	\caption{The BPT diagram for more than 35000 narrow line objects (contour in blue) and SDSS J1557. The solid purple and green lines (left panel) represent the dividing lines reported by \citet{Ka03} and \citet{Ke01}, and the solid purple lines (right panel) represent the dividing lines reported by \citet{Ke06}.
	The solid and dashed black lines in the right panel show the dividing line between AGN and
	HII galaxies and the area for expected composite galaxies discussed in \citet{Zh20}.
	The blue contour represents the distribution of more than 35000 narrow emission-line galaxies objects in SDSS DR15, as discussed in \citet{Zh20}, and the red five-pointed star and error bars denote the flux ratios and uncertainties for SDSS~J1557.
	}
	\label{fig5}
\end{figure*}

On the basis of Table 1, the flux ratio of [N{\sc ii}] line to H$\alpha$ line is 0.61, the flux ratio of [O{\sc iii}] line to H$\beta$ line is 1.37, and the flux ratio of [S{\sc ii}] line to H$\alpha$ line is 0.41.
SDSS~J1557 is plotted in the Baldwin-Phillips-Terlevich (BPT) diagram, as shown in Figure \ref{fig5}.
The standard optical diagnostic diagram, [O{\sc iii}]/H$\beta$ versus [N{\sc ii}]/H$\alpha$, of SDSS~J1557 is shown in the left panel of Figure \ref{fig5}.
Meanwhile, the diagnostic diagram, [O{\sc iii}]/H$\beta$ versus [S{\sc ii}]/H$\alpha$, of SDSS~J1557 is shown in the right panel of Figure \ref{fig5}.
Considering the dividing lines in the BPT diagrams \citep{Ke01,Ka03,Ci100}, SDSS~J1557 is a composite galaxy, and
the nucleus activity of SDSS~J1557 may be caused by galaxy merger.
We also try to strengthen the conclusion of AGN activity through two additional methods.
On one hand, the $W_1-W_2=0.309$ of SDSS J1557 does not satisfy the relationship of AGN that $W_1-W_2>0.5$ with 90\% completeness as described in \citet{As18}.
But there are much uncertainty that many random searched broad emission line AGNs in SDSS DR16 do not satisfy the $W_1-W_2>0.5$ relationship.
On the other hand, the calculated radio-loudness (R) of SDSS J1557 is 0.113, which means that SDSS J1557 is radio-quiet.
Moreover, it is also difficult to identify central nucleus activity from high-energy band characteristics due to lacking data from the ROentgen SATellite (ROSAT).
Therefore, more efforts are necessary to identify AGN activity in SDSS J1557.
In addition, given the composite classification and spatial resolution constraint, it is impossible to say which core provides spectral contribution.

%It is generally accepted that the stellar mass of galaxies is built up in the `inside-out' mode \citep{Ch97,Va10,Pa22}.
%According to top-hat spherical collapse model, the innermost shells firstly collapse, and the gas, which has the lowest angular momentum, enters the galaxy earliest \citep{Fa80,Ga17}.
%The inside-out growth of galaxies results in negative age gradients and is confirmed by hydrodynamic simulations \citep{Ro08,Ta16,Ti16}.
%The inside-out mode is a basic physical mode for galaxy growth and the structural evolution, which is a necessary part of the whole evolution \citep{Wal21}.
In the local universe, it has been proven that the growth of black holes is closely associated with galaxy evolution \citep{Su20}.
Central massive black holes (BHs) are found in massive galaxies with stellar mass $\gtrsim 10^{10}{M_\odot}$ \citep{Ma98}, and there is a scaling relation between black hole mass and bulge stellar mass \citep{Yo15,Se21}.
The relation \citep{Sa11} is given by
\begin{equation}
\log(M_{\rm BH}/{M_\odot})={\alpha}+{\beta}{\times}[\log(M_{\rm \star,bul}/{M_\odot})-11],
\end{equation}
where $M_{\rm BH}$ is the black hole mass, and $M_{\rm \star,bul}$ represents the stellar mass of the bulge. Their best fitting is
$\alpha$=8.16$\pm$0.06 and $\beta$=0.79$\pm$0.08.
According to the result of pPXF, the total stellar mass of SDSS~J1557 is about $10^{12}{M_\odot}$.
In addition, the bulge-to-total stellar mass ratio (B/T) has been well studied \citep{Bl14,De20}, and bulge mass and total stellar mass are equivalent \citep{Re15}.
So the $M_{\rm BH}$ in SDSS~J1557 is about $10^{8.95}{M_\odot}$ when $M_{\rm \star,bul} \thicksim 10^{12}{M_\odot}$.
Considering that SDSS J1557 is likely an advanced merger, the result of best fitting stellar mass from pPXF should then be framed as having contributions from both parent galaxies and
the $M_{\rm BH}$ in SDSS~J1557 is then a combination of the black holes in the NE and SW cores.

There are several scaling relations between the property of host galaxy and its central BH mass. The most well-studied one is the relation between stellar velocity dispersion and central BH mass, which is known as $M_{\rm BH}-\sigma_{\ast}$ relation \citep{Fe00,Ge00,Gu09}.
The finding of this scaling relation is of great significance.
For relative high-mass galaxies, the $M_{\rm BH}-\sigma_{\ast}$ relation points to the feedback between host galaxy evolution and growth of its central BH \citep{Kr18}.
For low-mass galaxies, the $M_{\rm BH}-\sigma_{\ast}$ relation could give insight into the formation
mechanisms of BH seeds at high redshift, as well as into the BH growth efficiency in small galaxies \citep{Ba20}.
Many works show that AGNs share the same $M_{\rm BH}-\sigma_{\ast}$ relation as quiescent galaxies \citep{Gr04,Gr08,Ji22}, but some investigations suggest that this relation may change among different types of galaxies \citep{Gr09,Ra17,Do21}.
So it is a good opportunity to verify this relation for this dual core system.
\citet{Ko13} described the $M_{\rm BH}-\sigma_{\ast}$ relation (considering the individual errors) of classical bulges and elliptical galaxies as
\begin{equation}
	\begin{split}	
		\log(\frac{M_{\rm BH}}{10^{9}{M_\odot}})=-(0.501\pm0.049)+	
		\\
		(4.414\pm0.295)\times \log(\frac{\sigma_{\ast}}{200 \rm km s^{-1}}).
	\end{split}
\end{equation}
\citet{Ko13} do not claim that the $M_{\rm BH}-\sigma_{\ast}$ relation is a good application for a merged galaxy.
But some studies as discussed in \citet{Be15,Sa19,Be21} have reported that objects with pseudo-bulges, bars or signs of mergers 
do not tend to lie off the $M_{\rm BH}-\sigma_{\ast}$ relation due to no effect on the measured $\sigma$. 

\begin{figure}
	\centering\includegraphics[width = 8cm,height=5cm]{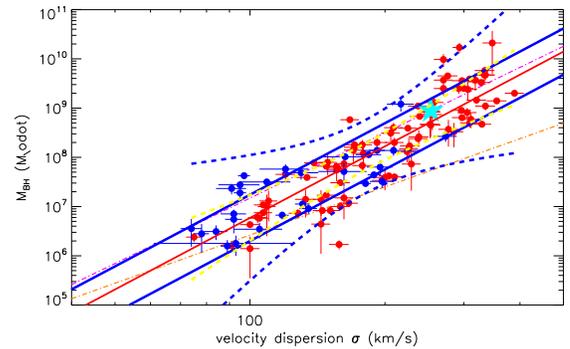}
	\caption{$M_{\rm BH}-\sigma_{\ast}$ relation. 
	The five-pointed star in cyan is SDSS J1557.	
	Solid circles in red and in blue show the values for the 89 quiescent galaxies and the 29 reverberation mapped AGNs, respectively.
	Dot-dashed lines in magenta and in orange represent the $M_{\rm BH}-\sigma_{\ast}$ reported in \citet{Ko13,Ho14}, respectively. The red solid line represents the best fitting with Linear Least-squares approximation. The dotted lines in yellow and blue show the confidence bands of 3$\sigma$ and 5$\sigma$ calculated by F-test, respectively.
	The blue solid line represents 1$\sigma$ scatter.}	
	\label{fig9}
\end{figure}

According to Figure \ref{fig5}, SDSS J1557 has probable nucleus activity, while the $M_{\rm BH}-\sigma_{\ast}$ relation determined in \citet{Ko13} is focous on quiescent galaxies. 
In order to show clearer information about the $M_{\rm BH}-\sigma_{\ast}$ relation of SDSS J1557, it is useful to explore the $M_{\rm BH}-\sigma_{\ast}$ relation to active galaxies.
Therefore, the public lts\_linefit code discribed in \citet{Ca13} is applied to re-determine the $M_{\rm BH}-\sigma_{\ast}$ relation, through the 89 quiescent galaxies and the 29 reverberation mapped AGNs, shown as the solid red line in Figure \ref{fig9}, and this relation can be described as:
\begin{equation}
	\begin{split}	
		\log(\frac{M_{\rm BH}}{10^{9}{M_\odot}})=-(0.78\pm0.49)+	
		\\
		(4.83\pm0.22)\times \log(\frac{\sigma_{\ast}}{200 \rm km s^{-1}}).
	\end{split}
\end{equation}
According to the pPXF technique determined stellar velocity dispersion about 253\kms in SDSS J1557,
obviously, the SDSS J1557 can well follow the $M_{\rm BH}-\sigma_{\ast}$ relation in equation (5) with 3$\sigma$ confidence level.
In addition, in order to test whether other $M_{\rm BH}-\sigma_{\ast}$ is applicable in SDSS J1557, the relations of classical bulges reported in \citet{Ko13} and pseudobulges reported in \citep{Ho14} are also shown by dot-dashed magenta line and dot-dashed orange line in Figure \ref{fig9}, respectively.
Therefore, it seems to be acceptable that the galaxy merger would not change the $M_{\rm BH}-\sigma_{\ast}$ relation.
Moreover, it is emphasized that the velocity dispersion and black hole mass analyzed above are contributed from the two cores.	

\begin{figure}
	\centering\includegraphics[width=8cm,height=8cm]{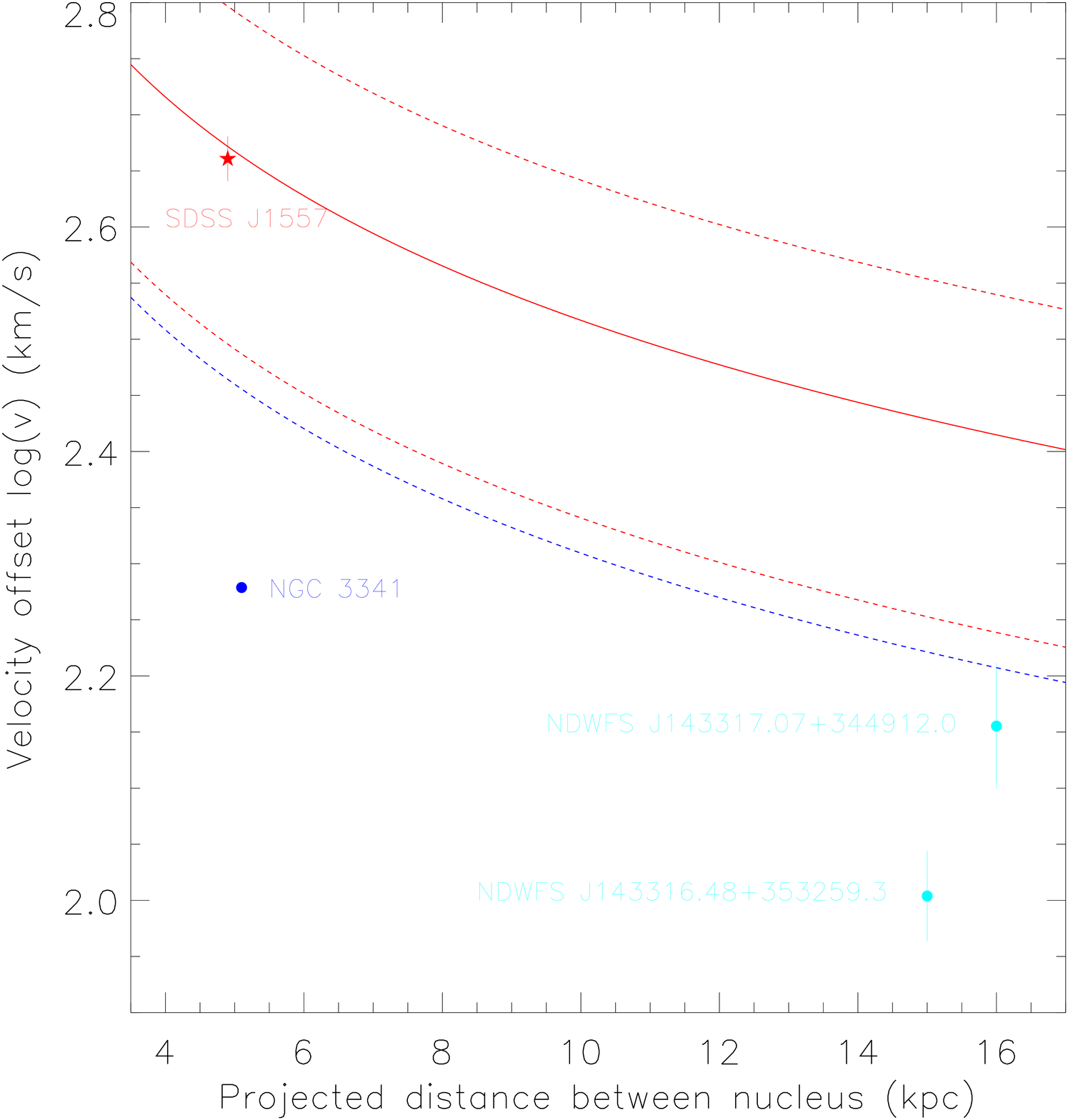}
	\caption{The relation between velocity offset and projected distance according to Table 2. The solid red line represents the mass ratio of two cores is 1:1. The dashed red lines represent the mass ratio of two cores is 2:1 and 1:2, respectively. The dashed blue line shows the relation between velocity offset and distance based on the mass reported in \citet{Ba08}.
	}
	\label{fig4}
\end{figure}

%In some cases, velocity offset of emission lines can be explained as the result of a strong galactic wind in the NLR of the AGN \citep{Ko08}.
%When galactic winds flow outwards, the enclosed mass increases, which will cause the galactic wind to slow down with distance from the galactic center.
There are two theoretical models which can be applied to explain
the velocity offset. In this paper, AGN-driven outflow model and dual core system are mainly discussed as follows.

In some cases, velocity offset of emission lines can be explained as the result of outflow.
For a centrally driven outflow, the velocity and electron density are maximal when closer to the
central engine, and would produce a stratified velocity structure \citep{Ko22}.
Based on the ionizations states, there are different origin and acceleration mechanisms between types of outflows
\citep{Cr03}.
Different combinations of thermal, magnetic, radiation, and shock driven forces can generate and accelerate winds and result in different results \citep{La21}.	
However, the emission lines listed in Table 1 show similar velocity offsets.
Due to the uncertainties of velocity offset in emission lines, it is reasonable that there are some differences in the velocity offset of different emission lines. 
Particularly, the H$\alpha$ and H$\beta$ emission lines are redshifted with respect to their absorption lines, which is contrary to the outflow model.	
So the model of AGN-driven outflow can not explain the significant velocity offsets well.

Another explanation for velocity offset is the dual core system.
The offset emission lines are caused by moving of black holes with respect to host galaxy.
As a result, the emission lines emitted in AGN NLR region may have velocity offset with respect to the absorption lines generated from host galaxy.
Apparently, the photometric image of SDSS~J1557 shows two bright nuclei (see Figure \ref{fig8}), and the nucleus activity might be the result of an ongoing merger.
In this case, SDSS~J1557 has a bulk motion relative to the host galaxy, so all the emission lines, in spite of ionization potential, would present  similar velocity offsets with respect to its host galaxy.
Since the spectrum is a composite of the two cores and only one set of discernible emission is seen, the emission could be originating from either of the cores, although it is impossible to determine which core is actually the AGN because of the spatial resolution limits.
At least one of the cores is indeed offset from the underlying host galaxy.
It would be nice for possible follow-up observations that could potentially resolve the separate spectra of the two cores (e.g., Keck long-slit spectroscopy) in the future work.
Therefore, large velocity offset between emission and absorption lines might be a good indicator for dual core system in SDSS~J1557.

Considering the simplest model of rotation of binaries, regardless of eccentricity, the relation of circular motion is that:
\begin{equation}
	v_1=\sqrt{\frac{Gm_{2}r_{1}}{L^2}},
\end{equation}
where $v_1$ and $r_1$ are velocity and radius of circular motion of one source, respectively. $L$ is the distance between two sources, and $m_2$ is the mass of another source.
It can be generally accepted that the velocity with respect to center will increase as two objects get closer.
Three cases of dual core systems with known velocity offset were reported in other references. Table 2 and Figure \ref{fig4} exhibit the velocity offset and the projected distance between two cores.
Based on the bulge mass estimate and equation (5), the relation between velocity offset and real distance can be given in Figure \ref{fig4}, assuming the mass ratio is 1:1.
The source in our paper, SDSS~J1557, can well fit the relation between velocity offset and distance, denoted by the solid red line. 
\citet{Co13} reported the information about distance and velocity offset between two cores, but the estimation of galaxy mass is lacking.  
In strict,  NGC~3341 is found to be a triple core system, and the stellar masses of two major cores are $10^{11}{M_\odot}$ and $4\times 10^{9}{M_\odot}$, respectively \citep{Ba08}. Regardless of the minor core, NGC~3341 is found to have a rather lower velocity offset than the prediction of binary model.
%The large difference may be due to: (1) The distance provided here is projected distance rather than real distance; (2) The mass of difference sources are on different magnitude.
%According to the calculation in \citet{Co13},the relation of velocity offset and real distance is shown as the solid red line in Figure \ref{fig4}. 
It implies that the projected distance between two major cores in NGC 3341 is rather smaller than real distance.

%So it is wonderful that the projected distance of SDSS~J1557 might be the real distance if the two cores have similar mass.

As reported in \citet{De14}, SDSS J150243.09+1111557.3 is a triple black hole system, and has double-peaked [O~{\sc iii}] emission lines.
Its closest pair is separated by about 140 parsecs, and the distance of 7.4 kpc between J1502P and J1502S is similar to the distance of 4.9 kpc between two cores in this paper.
It is interesting that \citet{Bo09} reported a galaxy, SDSS J092712.65+294344.0, showing two sets of broad emission line
and one set of absorption line. 
Two black holes with a separation of 0.1 pc have masses of $10^{7.3}{M_\odot}$ and $10^{8.9}{M_\odot}$, which have similar total mass of black holes to that in SDSS~J1557, and the velocity offset between two sets of broad emission line is 3500 \kms.	
These two black holes are orbiting each other with their BLRs, just like the two cores of SDSS~J1557 are orbiting each other with their bulges.
It is easily understood that the closer separation between
two cores in SDSS J092712.65+294344.0 leads to a higher velocity offset.

\section{Future Applications}
Obviously, the velocity offset of the emission lines (i.e., a large deviation from the central wavelength at the rest frame) usually predicts the presence of unique dynamic characteristics \citep{Ma16}.
%In this paper, a galaxy with a velocity offset is considered as a BBH system.
The velocity offset of emission lines might be an effective indicator to find dual core system.
Based on the 16th Data Release (DR16) of the Sloan Digital Sky Survey (SDSS) \citep{Ah20}, a sample of dozens galaxies with large velocity offset of emission lines can be selected.
Though it is rather time-consuming to fit the spectrum with pPXF code and to search the dual cores with photometric images, an investigation based on large sample of dual core system will be carried out in the near future.
Based on the sample of dozens galaxies with large velocity offset of emission lines, it is hopeful to test effectiveness of the criterion of large velocity offset of emission lines.

In addition, double-peaked profile in narrow emission lines caused by inspiralling SMBHs during a galaxy merger, is also a powerful observational tool for identifying dual AGN system.
We prepare to collect a new and large sample of galaxies with dual cores and possible double-peaked narrow emission-lines from SDSS DR16, because the current reported objects with double-peaked emission lines are from SDSS DR7 or DR8.
Based on the properties of velocity offset estimated by projected separation, it is of interest to test whether the double-peaked emission lines are caused by dual AGNs in the near future.

\section{Summaries and Conclusions}
In this paper, a dual core system with significant velocity offset between emission and absorption lines is found in the galaxy SDSS~J1557. Main conclusions are as follows:
\begin{itemize}
\item The starlight from host galaxy is determined by pPXF code with 636 SSPs. Stellar mass of SDSS J1557 is about $10^{12}{M_\odot}$, and stellar velocity dispersion is about 253 \kms. 
\item Based on the absorption lines modeled by SSP method and the emission lines modeled by Gaussian functions, SDSS~J1557 is found to have a significant velocity offset of 458$\pm$13 \kms, defined by H$\alpha$, between the frames of emission and absorption lines. Various emission lines have similar velocity offsets.
\item The mass of two black holes of SDSS~J1557 is estimated about $10^{8.95}{M_\odot}$. SDSS~J1557 can well fit the $M_{\rm BH}-\sigma_{\ast}$ relation of bulges, indicating that dual core system could not change this relation.
\item The model of AGN-driven outflow fails in interpreting the similar velocity offsets for various emission lines and the systematic redshift of emission lines relative to absorption lines. 
\item At least one nucleus is optically active, while the other is quiescent given the observed lack of two sets of distinct emission lines.	 
\item The large velocity offset between emission lines and absorption lines might be an effective indicator of dual core system, and the investigation based on a large sample of dozens galaxies with dual cores will be carried out in the near future.
\end{itemize}

\section*{Acknowledgements}
%{\color{red}Zhang gratefully acknowledge the
%anonymous referee for giving us constructive comments
%and suggestions to greatly improve the paper.}
%{Zheng, Zhang and Yuan gratefully acknowledge the anonymous referee for reading our paper carefully and patiently. }
This work is supported by the National Natural Science Foundation of China (Nos. 11873032, 12173020, 12273013). We have made use of the data from SDSS DR16.
The pPXF code website is (\url{http://www-astro.physics.ox.ac.uk/~mxc/idl/#ppxf}).
The MPFIT website is (\url{http://cow.physics.wisc.edu/~craigm/idl/idl.html}).
The paper has made use of the data from the MILES (\url{http://miles.iac.es/}) developed for stellar population synthesis models.
The emcee package website is \url{https://emcee.readthedocs.io/en/stable/}.

This Letter has made use of the data from
the SDSS projects.
The SDSS DR16 website is (\url{http://skyserver.sdss.org/dr16/en/home.aspx}).
The SDSS-III web site is \url{http://www.sdss3.org/}. SDSS-III is managed
by the Astrophysical Research Consortium for the Participating Institutions of the SDSS-III
Collaboration.

The DESI Legacy Imaging Surveys consist of three individual and complementary projects: the Dark Energy Camera Legacy Survey (DECaLS), the Beijing-Arizona Sky Survey (BASS), and the Mayall z-band Legacy Survey (MzLS). DECaLS, BASS and MzLS together include data obtained, respectively, at the Blanco telescope, Cerro Tololo Inter-American Observatory, NSF’s NOIRLab; the Bok telescope, Steward Observatory, University of Arizona; and the Mayall telescope, Kitt Peak National Observatory, NOIRLab. NOIRLab is operated by the Association of Universities for Research in Astronomy (AURA) under a cooperative agreement with the National Science Foundation. 

%\iffalse
%\section*{Data Availability}
%The data underlying this article will be shared on reasonable request to the corresponding author (aexueguang@qq.com).
%\fi

\end{document}